\documentclass[12pt,preprint]{aastex}
\usepackage{epsfig}
\newcommand{\HII}{\mbox{H\,{\sc ii}}}

\newcommand{\htco}{H$_{2}$CO}

\newcommand{\kmps}{km s$^{-1}$}

\newcommand{\pcmcub}{\mbox{${\rm cm^{-3}}$}}

\slugcomment{}

\shorttitle{}
\shortauthors{}

\received{}		

\begin{document}

\title{Anomalous \htco\ Absorption Towards the Galactic Anticenter: \\
A Blind Search for Dense Molecular Clouds \\[0.2in]
}

\author{M\'onica Ivette Rodr\'{\i}guez$^{1,2}$\\
Ronald J. Allen$^{1}$ \\
Laurent Loinard$^{2}$ \\
Tommy Wiklind$^{1,3}$ \\[0.2in]}

\affil{$^{1}$Space Telescope Science Institute, 3700 San Martin Drive,
Baltimore, MD 21218, USA \\ monica, rjallen, wiklind@stsci.edu}

\affil{$^{2}$Centro de Radiostronom\'{\i}a y
Astrof\'{\i}sica, Universidad Nacional Aut\'onoma de M\'exico,
Apartado Postal 72--3 (Xangari), 58089 Morelia, Michoac\'an,
M\'exico\\ m.rodriguez, l.loinard@astrosmo.unam.mx}

\affil{$^{3}$Affiliated with the Space Sciences Department of the European Space Agency}

\begin{abstract}
We have carried out a blind search in the general direction of the Galactic Anticenter for
absorption of the Cosmic Microwave Background (CMB) radiation near 4.83 GHz by
molecular clouds containing
gaseous ortho-formaldehyde (\htco). The observations were done using the 25-m radio telescope
at Onsala in Sweden, and covered strips in Galactic latitude
$-1^{\circ} \leq b \leq +1^{\circ}$
at several longitudes in the region $170^{\circ} \leq l \leq 190^{\circ}$. Spectra were
obtained in these strips with a grid spacing corresponding to the telescope resolution of $10'$.
We have detected \htco\ CMB absorption at $\approx 10\%$ of the survey pointings.
This detection rate is likely to increase with further improvements in sensitivity,
and may become comparable to the detection rate expected from a blind CO survey with
a corresponding sensitivity limit. We have mapped
some of these detections in more detail and compared the \htco\ absorption to existing maps of
CO(1-0) emission in the same regions. There appears to be a rough correlation between the
velocity-integrated line strength of the CO(1-0) emission and that of the \htco\ absorption.
However, the scatter in this correlation is significantly larger than the measurement errors,
indicating differences of detail at and below the linear resolution of our observations
($\approx 4 - 9$ pc). Although these two tracers are expected to have similar excitation
requirements on the microscopic level characteristic of warm, T$_K > 10$K, dense,
$10^3 < n < 10^5$ cm$^{-3}$ condensations in molecular clouds, the CO(1-0) line is expected to
be optically thick, whereas the \htco\ line is not. This latter difference is likely to be
responsible for a significant part of the scatter in the correlation we have found.
\end{abstract}


\section{Introduction}
\label{sec:introduction}

Following on the discovery of the anomalous absorption of the Cosmic Microwave Background (CMB)
at 4.83 GHz by gas-phase ortho-formaldehyde (\htco) molecules in a few nearby Galactic dark nebula
\citep{pzb69}, several
surveys were carried out in hopes of establishing the general Galactic distribution of distant
dusty molecular clouds containing gaseous \htco. \citet{gr71} used the NRAO 140-foot telescope
(beam FWHM $\approx 6'$, system temperature T$_S \approx 85$ K) to survey 30 positions spread
out along the Galactic plane near $b = 0^{\circ}$ in the range
$2.0^{\circ} \leq l \leq 251.1^{\circ}$ and chosen to be free
of radio continuum emission. With the exception of the one position near to the Galactic center
(which did indeed appear to have continuum emission in the general area), no detections of CMB
absorption could be registered, with a typical peak limits (5$\sigma$) of 0.07 K in a velocity
channel of width 1.6 \kmps. Gordon \& Roberts
concluded either that the excitation temperature of the general Galactic distribution of \htco\
must be close to the brightness temperature of the CMB (now known to be 2.73 K), or that the dust
clouds harboring the \htco\ must be much smaller than the telescope beam. This disappointing
result was corroborated with additional observations by \citet{gh73} using
the 25-m radio telescope of the Onsala Space Observatory (OSO) in Sweden (FWHM $\approx 10.5'$,
T$_S \leq 55$ K) in order to map three $1^{\circ} \times 1^{\circ}$ fields in the Galactic
plane at $l = 48^{\circ}, 70^{\circ}$, and $110^{\circ}$. Spectra were obtained in each of
these fields on a grid with separation $10'$. No emission or absorption was found, with a
typical peak limit of 0.15 K in a 0.62 \kmps\ channel.

The first large-scale survey for \htco\ along the Galactic plane was carried out by \citet{f79}
using the Jodrell Bank Mark II radio telescope (FWHM $9.6' \times 10.3'$, T$_S \approx 70$ K).
Observations were made at $b = 0^{\circ}$ every $2^{\circ}$ of Galactic longitude in the range
$8^{\circ} \leq l \leq 60^{\circ}$ and every $1^{\circ}$ in the range
$14^{\circ} \leq l \leq 36^{\circ}$. These observations were successful in recording \htco\
absorption in the inner Galaxy, and some approximate information on the spatial distribution
was obtained. However, the signal dropped to undetectable levels beyond $l \gtrsim 50^{\circ}$,
and no observations were attempted in the outer Galaxy. In fact, at no position did the
absorption-line profile depth exceed the observed continuum temperature. One can therefore
safely conclude that what was being measured was not the anomalous CMB absorption (which was
still apparently too weak), but rather absorption of the Galactic background radio
radiation, which is strongest in the inner Galaxy.

Although gaseous \htco\ may be nearly absent in parts of the ISM because it is dissociated
or frozen out on grains, these early surveys provided an indication that the physical
conditions under which detectable anomalous \htco\ absorption occurs may not be common
in the Galaxy. Observers turned their attention to other, more easily detected
molecules, notably CO, and we now have extensive surveys of the CO(1-0) line over large sections
of the Galactic plane \citep{duc87, c91, dht01}.
These surveys, along with many detailed studies of specific molecular
clouds in a wide range of molecular tracers, have all contributed to a much more complete (and
much more complicated) view of physical conditions in the cool molecular ISM.

An explanation for the anomalous CMB absorption by \htco\ was first suggested by \cite{tc69}
using a classical calculation for collisional excitation. Subsequent work, especially by Evans
and his collaborators \citep{e75, ezms75}, confirmed this result using quantum mechanical
calculations and observations. The collisional pumping mechanism is more effective at high
collision rates, so in general the absorption is strongest at higher densities and temperatures.
However, the calculations by \cite{e75} showed that the mechanism would still be effective at
rather low temperatures, below about 10K. More precise quantum mechanical calculations
reported by \cite{glmg75} suggested a smaller effect at very low kinetic temperatures, but
both methods involved approximations. This leaves open the possibility that high-density,
cold clumps of molecular gas in the ISM may be detectable in \htco\ absorption, and has
provided part of the motivation for our observing program.

Our approach is to carry out long integrations at a set of blindly-selected positions in the
general direction of the Galactic Anticenter. For these observations we have again used the
25-m OSO radio telescope at Onsala, the same telescope used by Gordon \& H\"oglund more
than 33 years ago in their failed attempt to detect the general \htco\ CMB absorption. Our
present success is due entirely to the availability of more sensitive receivers and to generous
allocations of observing time on this telescope. Our survey has two main features: First,
it is a ``blind'' survey; we purposely avoided using maps of any other ISM tracer to construct
the observing program. Second, we chose to observe in the Galactic anticenter region, i.e.
in the general direction of the outer Galaxy. The Galactic nonthermal background at 6 cm
is exceedingly faint in this direction, so we can be fairly confident that any absorption we
might detect is indeed anomalous CMB absorption. We will return to this point later.
Also, the velocity gradient owing to Galactic rotation is small in the anticenter direction,
so we might hope for some degree of ``bunching'' of absorption features, enhancing the
probability of detection.

\section{Observations and results}
\label{sec:observations}

The observations were obtained during two sessions, the first in September-October 2004, and
the second in May 2005, with the 25m Onsala radio telescope.  At 6 cm, the angular resolution
of the telescope is FWHM $\approx 10'$,
and the pointing precision is better than $20''$. The local oscillator was operated in
frequency-switching mode with a ``throw'' of 0.8 MHz, and both (circular) polarizations
of the incoming signal were recorded simultaneously in two independent units of a digital
autocorrelator spectrometer. Each of these units provided 800 channels spaced by 4 kHz.
At the frequency of the 1$_{11}$-1$_{10}$ line of ortho-\htco\ (4829.660 MHz), this setup
provided a total bandwidth of 3.2 MHz = 199 km s$^{-1}$ and a (twice-hanning-smoothed) velocity
resolution of 16 kHz = 0.99 km s$^{-1}$. Daily observations of Cas A were made
in order to check the overall performance of the receiver. The system temperature during
our sessions varied from 33 to 36 K. The off-line data reduction was done with the
CLASS/GILDAS software system \citep{gf89}, and involved only the subtraction of (flat)
baselines from individual integrations and the averaging of all spectra taken at the same
pointing position. The total integration time at each pointing position was $\approx 2$
hours, yielding a final typical rms noise level of 0.0035 K (T$^*_A$) and a 5$\sigma$
detection limit of 0.0175 K in each $\approx 1$ \kmps\ channel.

\subsection{The blind search}
\label{subsec:blindsearch}

For our blind search, integrations were made every $10'$ in 11 strips perpendicular to the
Galactic plane from $-1^{\circ}$ to $+1^{\circ}$ at intervals of $2^{\circ}$ in Galactic
longitude from $170^{\circ}$ to $190^{\circ}$.  The spectrometer was centered at a slightly
different systemic velocity for each strip (cf. Table \ref{table:velocity}), following the
expected run of radial velocity according to the ``standard'' model of rotation in the outer
Galaxy using HI data from \citet{h97}.

\begin{deluxetable}{cccccccccccc}
\tablecaption{Spectrometer systemic velocity used at each longitude \label{table:velocity}}
\tablewidth{0pt}
\tablehead{}
\startdata
Longitude (deg.) & 170 & 172 & 174 & 176 & 178 & 180 & 182 & 184 & 186 & 188 & 190 \\
Velocity (km s$^{-1}$) & -12 & -9.7 & -8.5 & -7.3 & -5.4 & -3.1 & -1.0 & 1.4 & 3.1 & 4.3 & 6.2
\enddata
\end{deluxetable}

\begin{figure*}[ht!]
\plotone{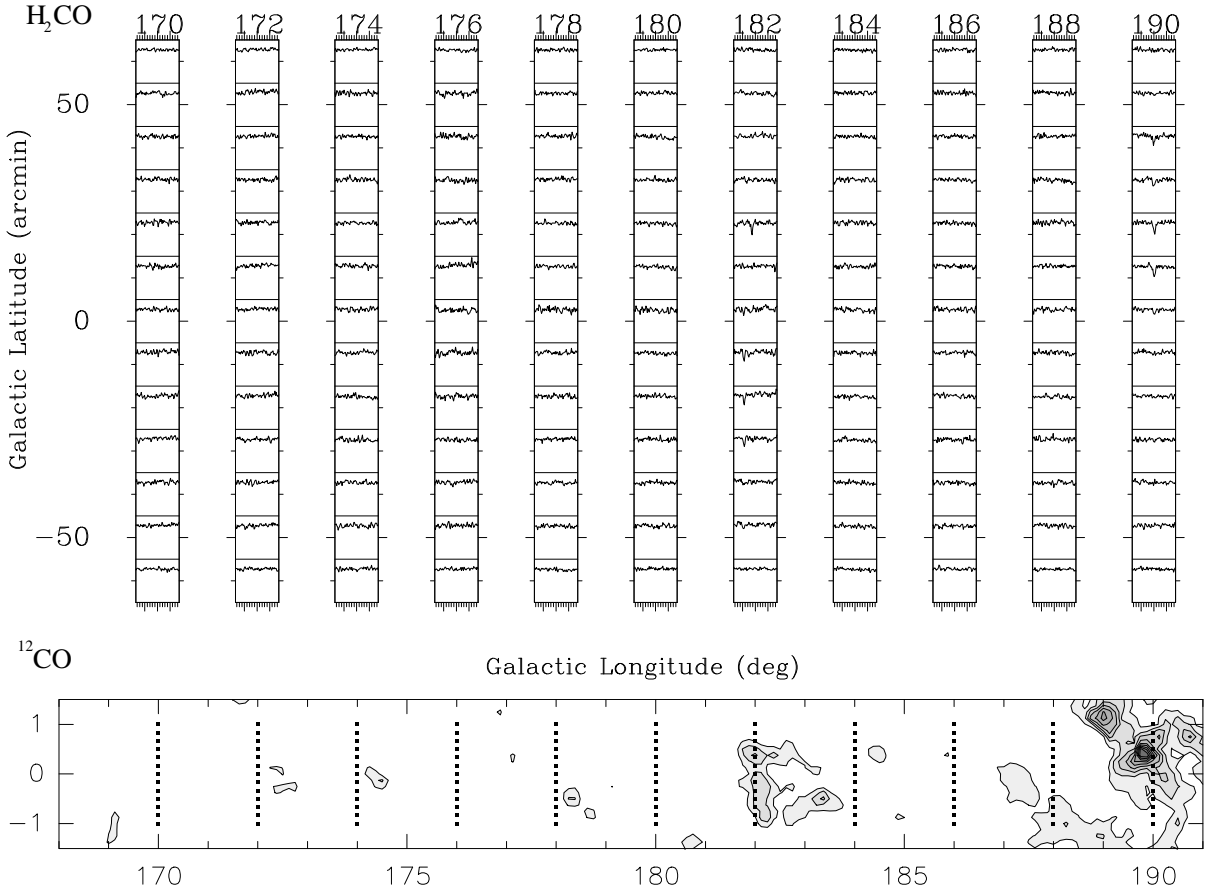}
\caption{Mosaic of \htco\ spectra at 143 positions towards the Galactic anticenter.
The Galactic longitude of each strip is given at the top of the figure, and the Galactic
latitude on the left side. The velocity scale along the bottom of each longitude strip
is centered at the values given in Table \ref{table:velocity}; the short tick marks indicate
increments of 1 km s$^{-1}$ and the longer ticks are at intervals of 5 km s$^{-1}$. The lower
panel shows the survey positions superposed on a map of the CO emission in the area, see
text for details.
\label{fig:blindsearch}}
\end{figure*}

The final (averaged) spectra are shown as a mosaic in Figure \ref{fig:blindsearch}.
There is clear evidence for absorption in 10 -- 12 positions, and hints of absorption in
several more.  It is evident that, with our increased level of sensitivity, we have
successfully detected \htco\ absorption at roughly 10\% of the total of 143 positions
observed. It is also interesting to note that we have not recorded any \htco\ emission;
this may indicate that very dense gas with $n \gtrsim 10^5$ (see \S \ref{sec:discussion})
is rare, with a low area filling factor, and/or the gas is much colder than about 10 K. For
reference, Figure \ref{fig:blindsearch} also shows the disposition of our survey positions
with respect to the CO emission obtained from the survey by \cite{dht01}. We will return
to this comparison later, but we wish to emphasize here that our choice of \htco\
survey positions took no prior account of the distribution of CO emission. Nevertheless, it
is clear from this Figure that the \htco\ absorption is seen most strongly in regions of
strong CO emission. Figure \ref{fig:blindsearch} also shows regions of faint CO emission
without corresponding \htco\ absorption; for instance several survey points at $l = 188^{\circ}$
are located in a region of faint CO emission. The absence of \htco\ absorption at such positions
is likely to be a reflection of the sensitivity limit of our observations. 

\subsection{Mapping observations}
\label{subsec:mapping}

The presence of two regions of relatively strong absorption can be identified in
Figure \ref{fig:blindsearch}, the first at $l \approx 182^{\circ}$, and the second at
$l \approx 190^{\circ}$. We have mapped these two regions in more detail in order
to examine the distribution of \htco\ absorption and to permit a point-by-point
comparison with the existing CO(1-0) emission surveys in this region of the Galaxy. The
receiver settings for these two maps were identical to those used for the blind survey,
and the total integration times per pointing position were similarly long.

\begin{figure*}[ht!]
\plotone{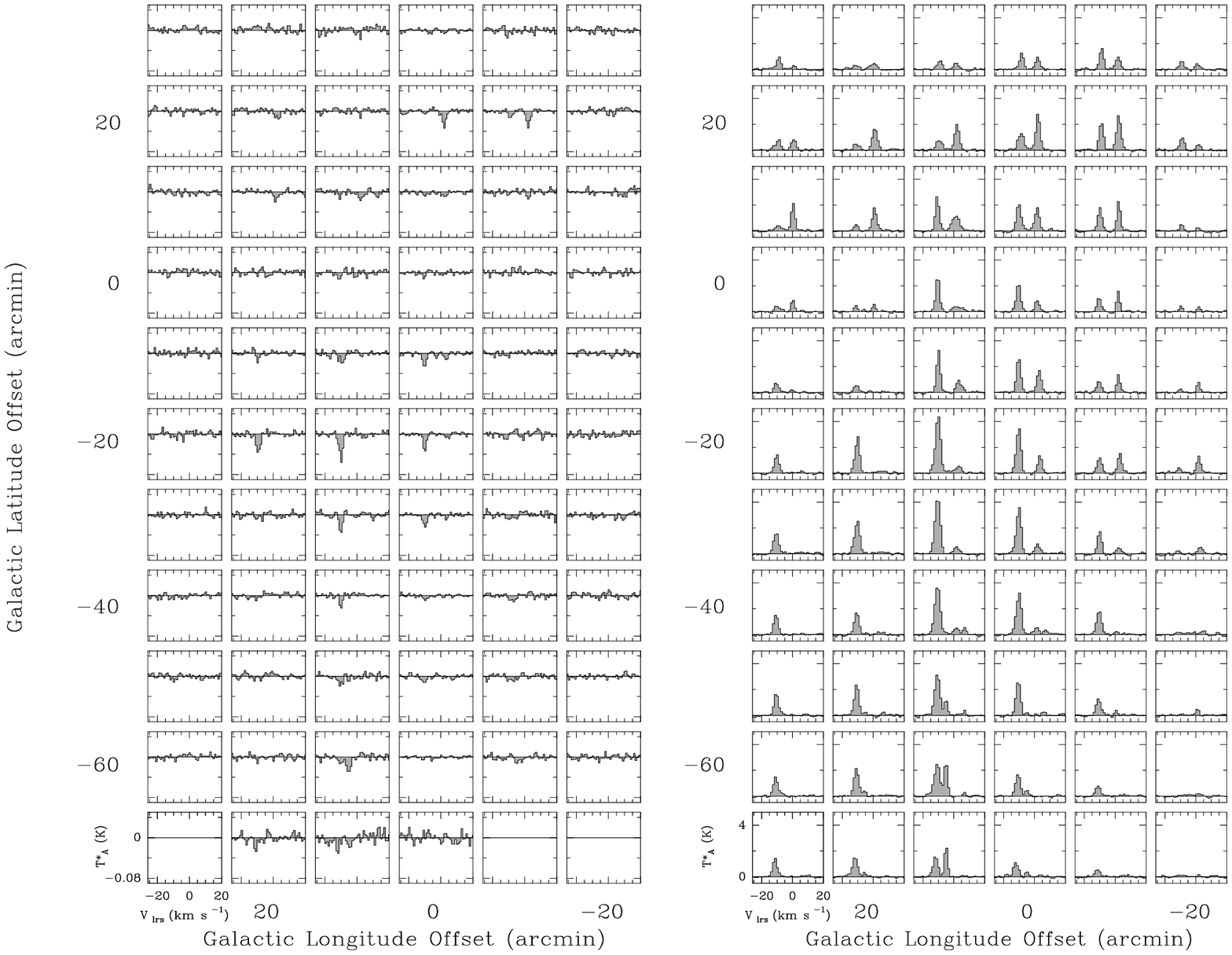}
\caption{a.(left panel) Mosaic of \htco\ spectra observed in the direction $l = 182^{\circ}$,
$b = 0^{\circ}$. b.(right panel) Observations of CO(1-0) at corresponding positions, from the
Columbia survey, smoothed to $10'$ (see text). 
\label{fig:firstmosaic}}
\end{figure*}

\begin{figure*}[ht!]
\plotone{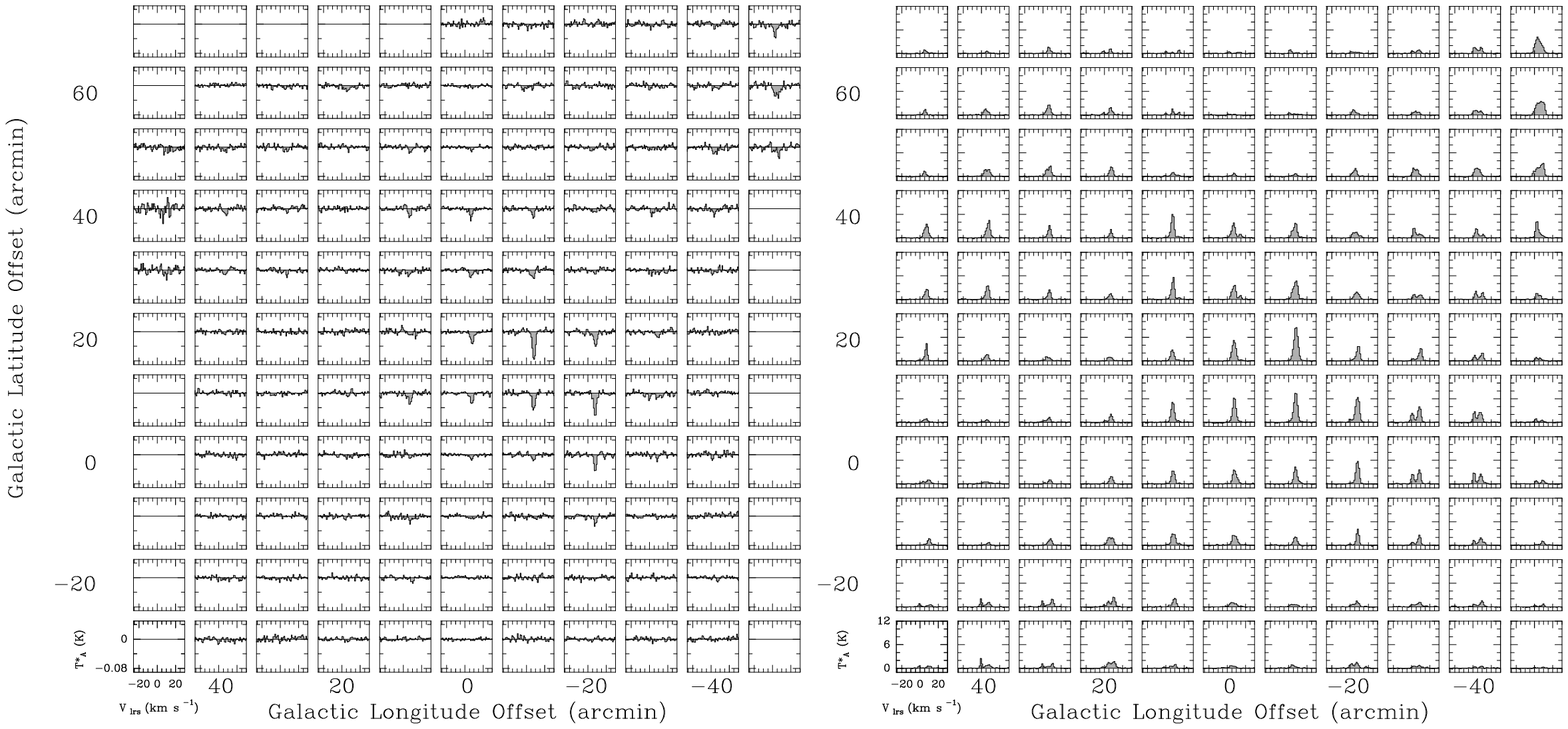}
\caption{a.(left panel) Mosaic of \htco\ spectra observed in the direction $l = 190^{\circ}$,
$b = 0^{\circ}$. b.(right panel) Observations of CO(1-0) at corresponding positions, from the
Columbia survey, smoothed to $10'$ (see text).
\label{fig:secondmosaic}}
\end{figure*}

Figure \ref{fig:firstmosaic}a (left panel) shows profiles at 63 of 66 positions on a grid
centered (position 0,0) at $l = 182^{\circ}$, $b = 0^{\circ}$, with an interval of $10'$. Similarly,
Figure \ref{fig:secondmosaic}a (left panel) shows profiles at 101 of 121 positions on a grid
centered (position 0,0) at $l = 190^{\circ}$, $b = 0^{\circ}$, also with an interval of $10'$.

\section{Analysis}
\label{sec:analysis}

\subsection{The nature of the \htco\ absorption}

\begin{figure*}[ht!]
\plotone{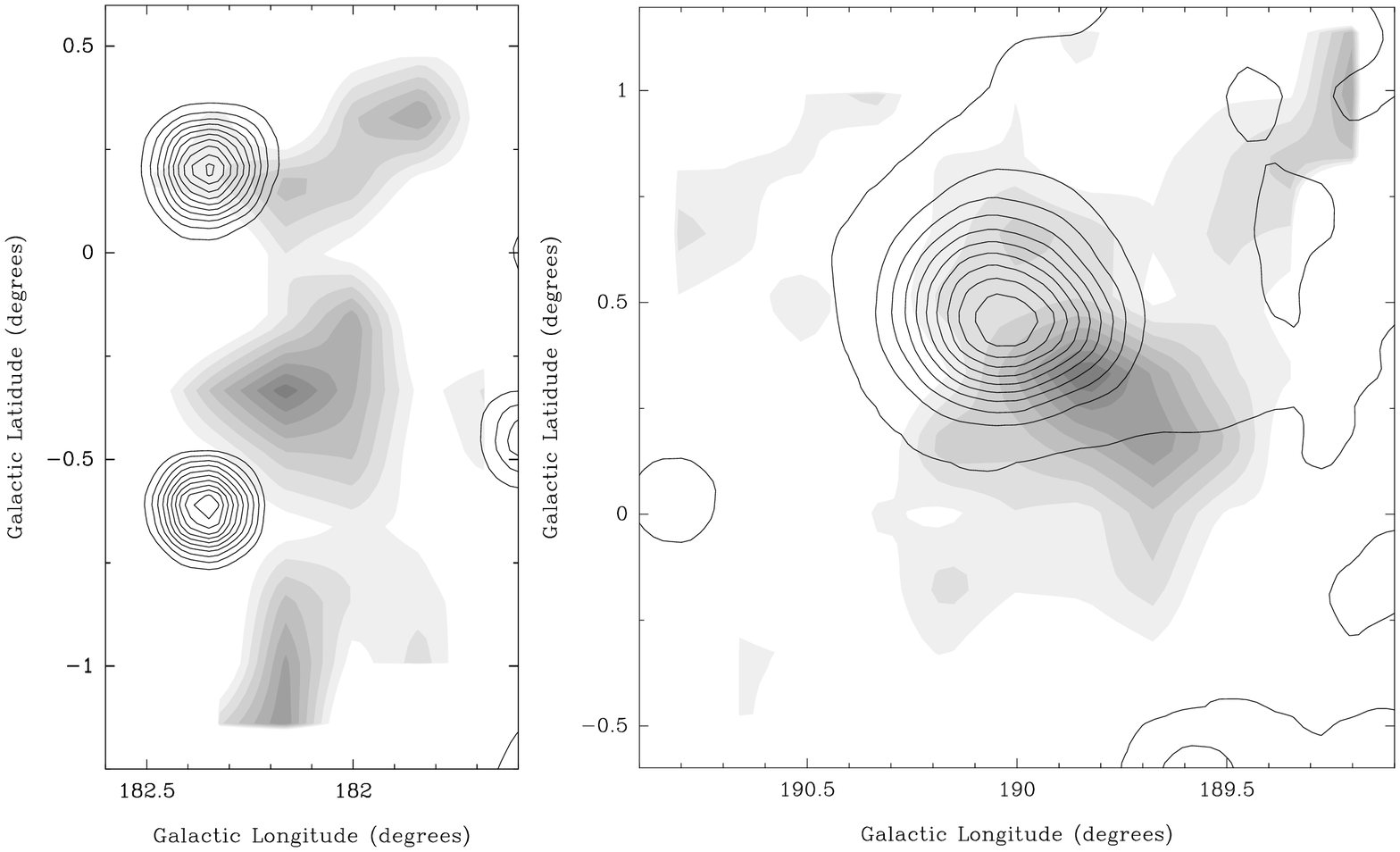}
\caption{The location of the 21-cm radio continuum emission (contours, Effelsberg 100-m
telescope, FWHM $9.4'$) overlaid on our \htco\ observations (grey scale, Onsala 25-m
telescope, FWHM $\approx 10'$). a.(left panel) detailed survey area near $l = 182^{\circ}$,
$b = 0^{\circ}$. b.(right panel) detailed survey area near $l = 190^{\circ}$, $b = 0^{\circ}$.
\label{fig:21cm}}
\end{figure*}

How can we be sure that the absorption we have recorded, both in our blind search and in our two
detailed maps, is actually absorption of the CMB? We have earlier argued that it is not
likely to be absorption of the Galactic nonthermal background, since this background is very
faint in the outer Galaxy at 6 cm, so we can confidently rule out the type of absorption recorded
in the inner Galaxy by \citet{f79}. But what about discrete continuum sources in the background,
perhaps distant \HII\ regions or radio galaxies? In order to examine this possibility, we have
retrieved radio continuum survey data at 21-cm from \citet{r97} covering the two regions we have
mapped in detail. Figure \ref{fig:21cm} shows the 21-cm continuum in these two regions observed
with the Effelsberg 100-m telescope at a resolution of $9.4'$ (contours) overlaid on our
\htco\ absorption data (grey scale, resolution $\approx 10'$). From the
general lack of correspondence of the radio continuum peaks with the \htco\ absorption maxima
we can safely conclude that it is indeed the CMB absorption we have recorded in \htco.
Furthermore, the continuum sources are either behind the \htco\ absorbers, or the area filling
factor of the absorbing clouds is small and the line of sight to the sources just misses the clouds.

\subsection{Statistics of the \htco\ absorption}

Our blind search has succeeded in recording anomalous \htco\
absorption at approximately 10\% of the survey positions. This is a significant improvement
over the earlier blind searches of \citet{gr71} and \citet{gh73}, which failed to record
any such absorption. Since we have no reason to believe that the regions chosen for the early
searches are in any way peculiar, we conclude that our success is likely due to the large
improvement in sensitivity of our observations. Our rms noise level is typically 0.0035 K
(T$^*_A$) for a 5$\sigma$ detection limit of 0.0175 K. When compared at the same velocity
resolution, this is an improvement of $\approx 5$
over the observations of \citet{gr71} and a factor $\approx 6.8$ over those of \citet{gh73}.
This result strongly suggests that the detection rate is simply sensitivity limited, and that 
it would increase with further increases in sensitivity. Future searches would benefit from
using larger telescope apertures with more collecting area; beam dilution would also be
reduced on the more distant dust clouds.

\subsection{Relation of \htco\ CMB absorption to CO(1-0) emission}

We have retrieved the CO(1-0) emission line data for our two mapped fields from the survey
by \citet{dht01}, smoothed that data slightly from its original $8.4'$ resolution to
the $10'$ resolution of our \htco\ observations, and extracted profiles at the same
positions as on our maps near $l = 182^{\circ}$, $b = 0^{\circ}$ and $l = 190^{\circ}$,
$b = 0^{\circ}$. The results are shown in the right panels of Figure \ref{fig:firstmosaic}
and Figure \ref{fig:secondmosaic}. A cursory inspection of these mosaics shows that
there is an overall general correlation between the two tracers, in particular, at every
position where we have detected \htco, CO(1-0) emission is also detected. The converse
is not always true; there are many positions where CO(1-0) is easily detected which show no
corresponding \htco\ above the noise. Although this may simply be a consequence of a lower
S/N for the \htco\ observations, a more detailed look at these profiles reveals that S/N is
not the whole story. In particular, the \htco\ absorption profiles are not merely scaled
versions of the CO(1-0) emission profiles. Consider the profiles on the mosaic of Figure
\ref{fig:secondmosaic} at $l = 190^{\circ}$, in particular those at (-20,+10), (-10,+10),
and (0,+10). In \htco\ this sequence of 3 profiles shows a uniform decrease in the depth
of the absorption from about -0.05 to -0.04 to -0.03 K, but in CO the emission profiles
peak at about 6, 7, and 6 K respectively.

\begin{figure*}[ht!]
\plotone{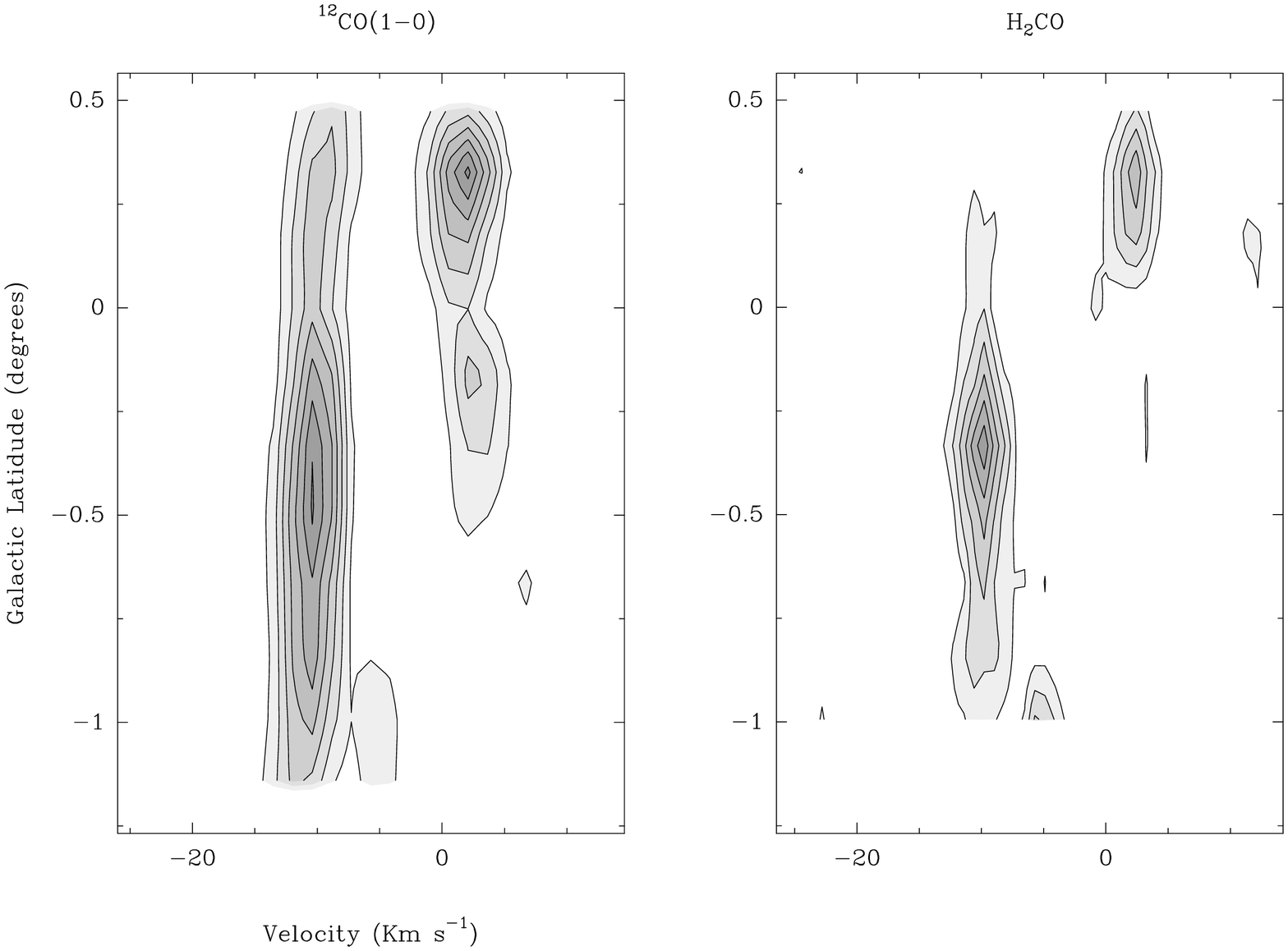}
\caption{a.(left panel) Latitude-velocity contour maps of $^{12}$CO emission and b.(right panel)
\htco\ absorption observed in the direction $l = 182^{\circ}$, $b = 0^{\circ}$. The data have
been integrated over the longitude range from $l = 181.7^{\circ}$ to $l = 182.5^{\circ}$ in
both panels. \label{fig:firstcomparison}}
\end{figure*}

\begin{figure*}[ht!]
\plotone{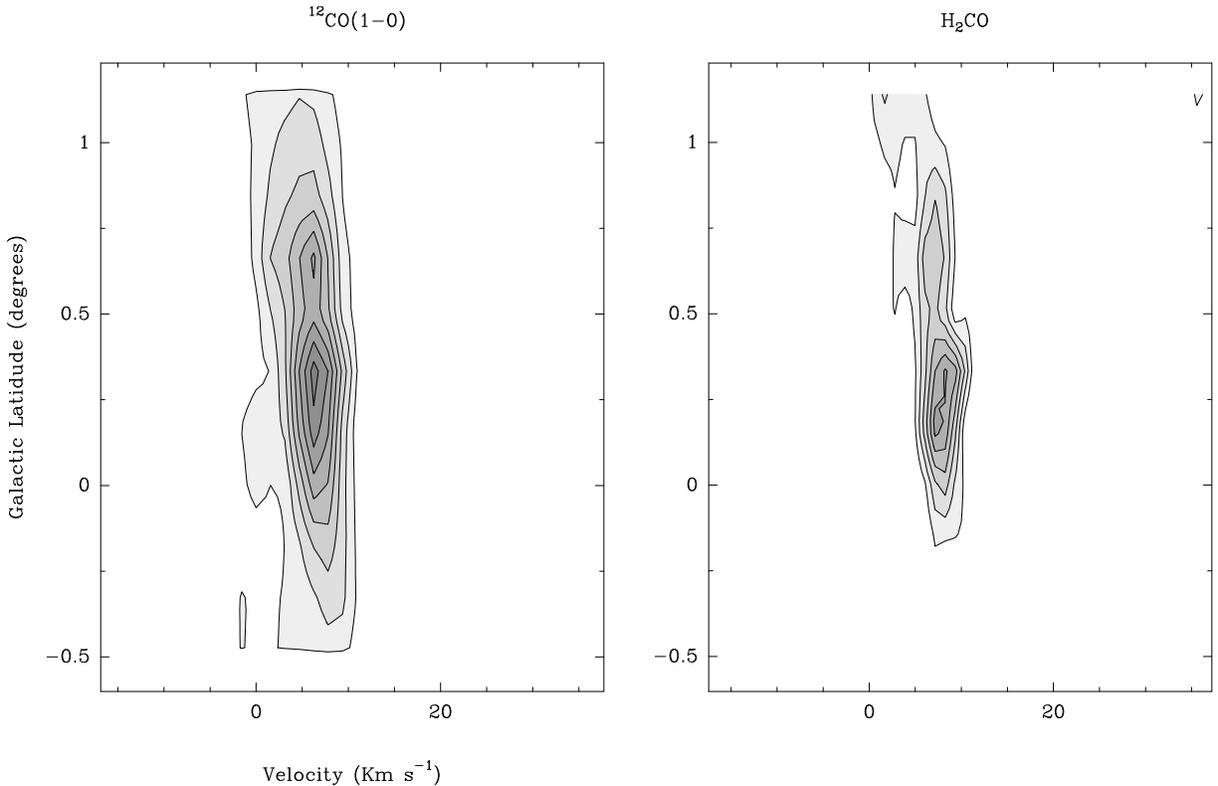}
\caption{a.(left panel) Latitude-velocity contour maps of $^{12}$CO emission and b.(right panel)
\htco\ absorption observed in the direction $l = 190^{\circ}$, $b = 0^{\circ}$. The data have
been integrated over the longitude range from $l = 189.2^{\circ}$ to $l = 190.8^{\circ}$ in
both panels. \label{fig:secondcomparison}}
\end{figure*}

Another way of showing the general correlation between CO and \htco\ is depicted in
Figures \ref{fig:firstcomparison} and \ref{fig:secondcomparison}, which show contour maps
of the two components in a latitude-velocity plot integrated over a small range in longitude.
Figure \ref{fig:firstcomparison} shows the two velocity components in this region, a weaker
component at $\approx +2.5$ \kmps, and a stronger component at $\approx -10$ \kmps. Figure
\ref{fig:secondcomparison} shows the data for the region centered at $l = 190^{\circ}$ in
the same latitude-velocity presentation. The principal component here is observed at
$\approx +8.5$ \kmps. These figures confirm the general similarity of the CO emission and the
\htco\ absorption. The \htco\ CMB absorption clearly generally traces the same molecular gas
seen in the more ubiquitous (and easier to observe) CO molecule.

A general correspondence of \htco\ and $^{12}$CO(1-0) was also noted by \citet{cmf83}
in their extensive mapping study of \htco\ and OH in the Orion region. Those authors
found a broad general agreement in that well-known star-forming region, but
concluded that the detailed agreement was poor. They found a better 
correspondence with $^{13}$CO(1-0), and concluded that the reason for this was that the
optically-thick $^{12}$CO(1-0) line was primarily tracing gas temperature, whereas the
$^{13}$CO(1-0) emission and the \htco\ absorption are both optically-thin lines that will
trace primarily the density. In general we agree with this interpretation, although as we will
discuss in more detail later the dependence of the \htco\ absorption on kinetic temperature
is also likely to be playing a role.

\begin{figure*}[ht!]
\plotone{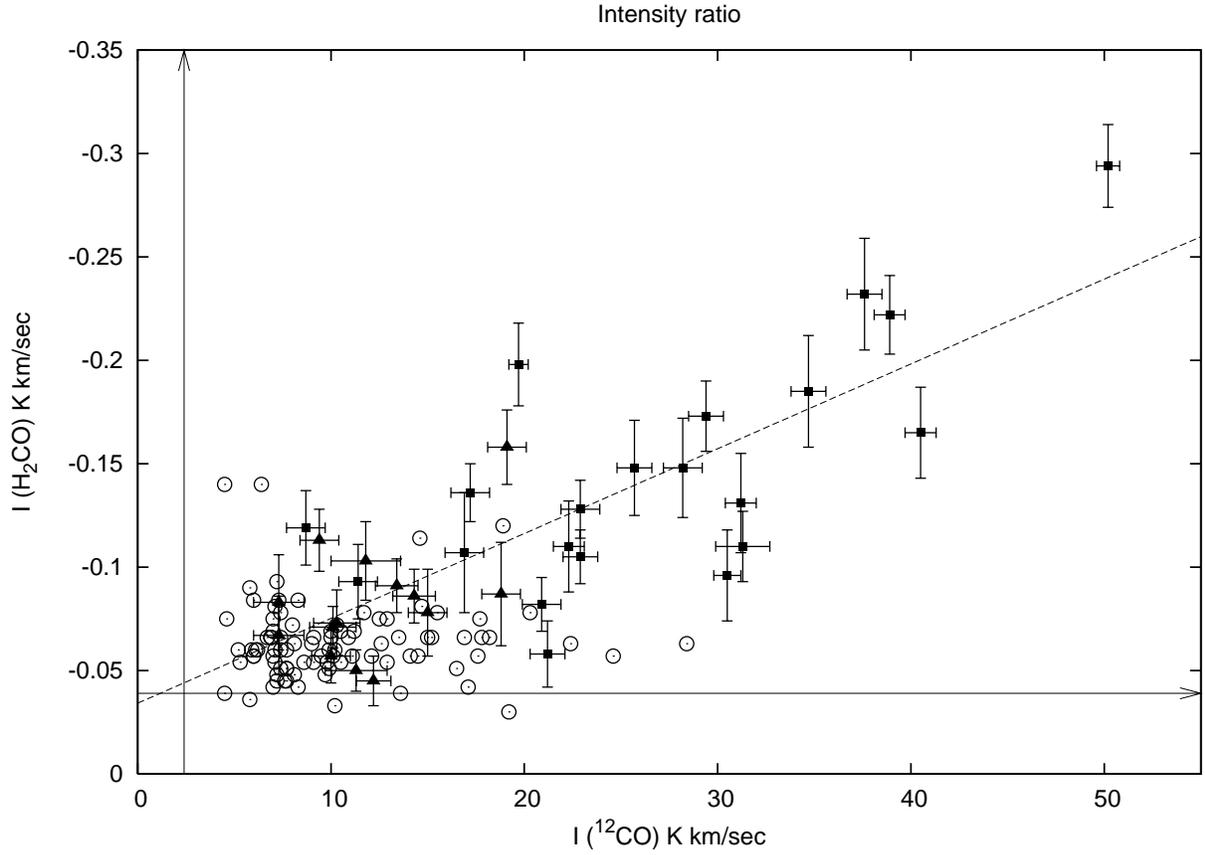}
\caption{ The correlation between the \htco\ absorption line intensity and the CO(1-0) emission
line intensity at corresponding points in the two fields surveyed in detail. The triangles
correspond to data from Table \ref{table:mom182}, the squares to data from Table
\ref{table:mom190}, and the open circles correspond to \htco\ upper limits (points where the CO was 
detected but \htco\ was not). The horizontal and vertical lines show the $\approx 3\sigma$ noise for 
the spectra with the lowest noise. The least squares fitted line is shown dashed. \textit{The data 
tables are shown in the electronic version of this paper.}}
\label{fig:intensityratio}
\end{figure*}

In order to study the relation between \htco\ absorption and CO(1-0) emission in more
detail, we have computed the moments of the profiles shown in
Figures \ref{fig:firstmosaic} and \ref{fig:secondmosaic}. The results are listed
in Tables \ref{table:mom182} and \ref{table:mom190} (\textit{see the electronic version of this paper}).
If two components are visible at a given pointing, the moments for each have been
computed separately; in many cases a second component was too weak to be identified in \htco.
Components that exceed the $3\sigma$ limit are indicated in column 6 of these
tables and their values plotted in the correlation diagram of Figure \ref{fig:intensityratio}.
Since the data in the two fields did not seem to show any different trends, we have included
the profile intensities from both fields in Figure \ref{fig:intensityratio}.

We have fitted the data in Figure \ref{fig:intensityratio} by least squares to a straight line;
the result is:

\begin{equation}
I(H_2CO) = a \times I(CO) + b
\end{equation}

\noindent yielding a value of $a = -0.0041 \pm 0.0005$ and $b = -0.034 \pm 0.013$.
While the fit is not bad, in several cases the data points lie significantly off the
best-fit line. We will return to this point in the Discussion below.

\subsection{Other sources in the two mapped fields}

In order to have some idea of the distances to our detections, we have searched the Catalog of
Star-Forming Regions in the Galaxy \citep{a02} for objects at
known distances which may be associated with them. Table \ref{table:IRAS} lists several IRAS
sources which are candidates, and their associated distances. These distances were obtained
from the references listed in the table, and were determined from optical (\HII\ regions) and
radio spectroscopy (CO(1-0) in molecular clouds) and using a model rotation curve of the outer
Galaxy.

\begin{deluxetable}{ccccc}
\tablecaption{IRAS sources possibly associated with CO and \htco\ emission in
Figures \ref{fig:iras180} and \ref{fig:iras190}. \label{table:IRAS}}
\tablewidth{0pt}
\tablehead{
\colhead{IRAS source} &
\colhead{Position (\textit{l,b})} &
\colhead{Associated object} & 
\colhead{Distance (kpc)} &
\colhead{References}
}
\startdata
IRAS 05490+2658             & 182.4, +0.3    &    S242   & 2.0 - 2.1  & 1, 2 \\
IRAS 05431+2629             & 182.1, -1.1     &          &    0.1 *       & 3    \\
IRAS 06067+2138             & 189.1, +1.1    &           &    0.9 *     & 3 \\
IRAS 06051+2041             & 189.7, +0.3    &           &    2.3 *     & 3 \\
IRAS 06055+2039             & 189.8, +0.3    &  Gem OB1, S252  & 1.5  - 2.9 * & 3, 4, 5  \\
IRAS 06063+2040             & 189.9, +0.5    &  Gem OB1, AFGL5183 & 1.5  - 2.8 *  & 3, 4, 6 \\
IRAS 06068+2030             & 190.1, +0.5    &  S252, AFGL5184 & 1.5  - 2.7 *& 1, 3, 6, 7 \\
IRAS 06061+2028             & 190.0, +0.3    &           &    2.3 *     & 3 \\
IRAS 06079+2007             & 190.5, +0.6    &           &    2.6 *    & 3 \\
\enddata
\tablecomments{
In all the cases the distances are taken from the published literature. \\
These distances are determined from spectroscopy of the exciting stars of the
associated \HII\ regions. \\
** Kinematic distances using $^{12}$CO data and a model rotation curve of the outer Galaxy.
}
\tablerefs{(1) \cite{b82}, (2) \citet{css95}, (3) \citet{wb89}, (4) \citet{h78}, 
(5) \citet{sdh90}, (6) \citet{shd88}, (7) \citet{mjf79}.  
}
\end{deluxetable}

\begin{figure*}[ht!]
\plotone{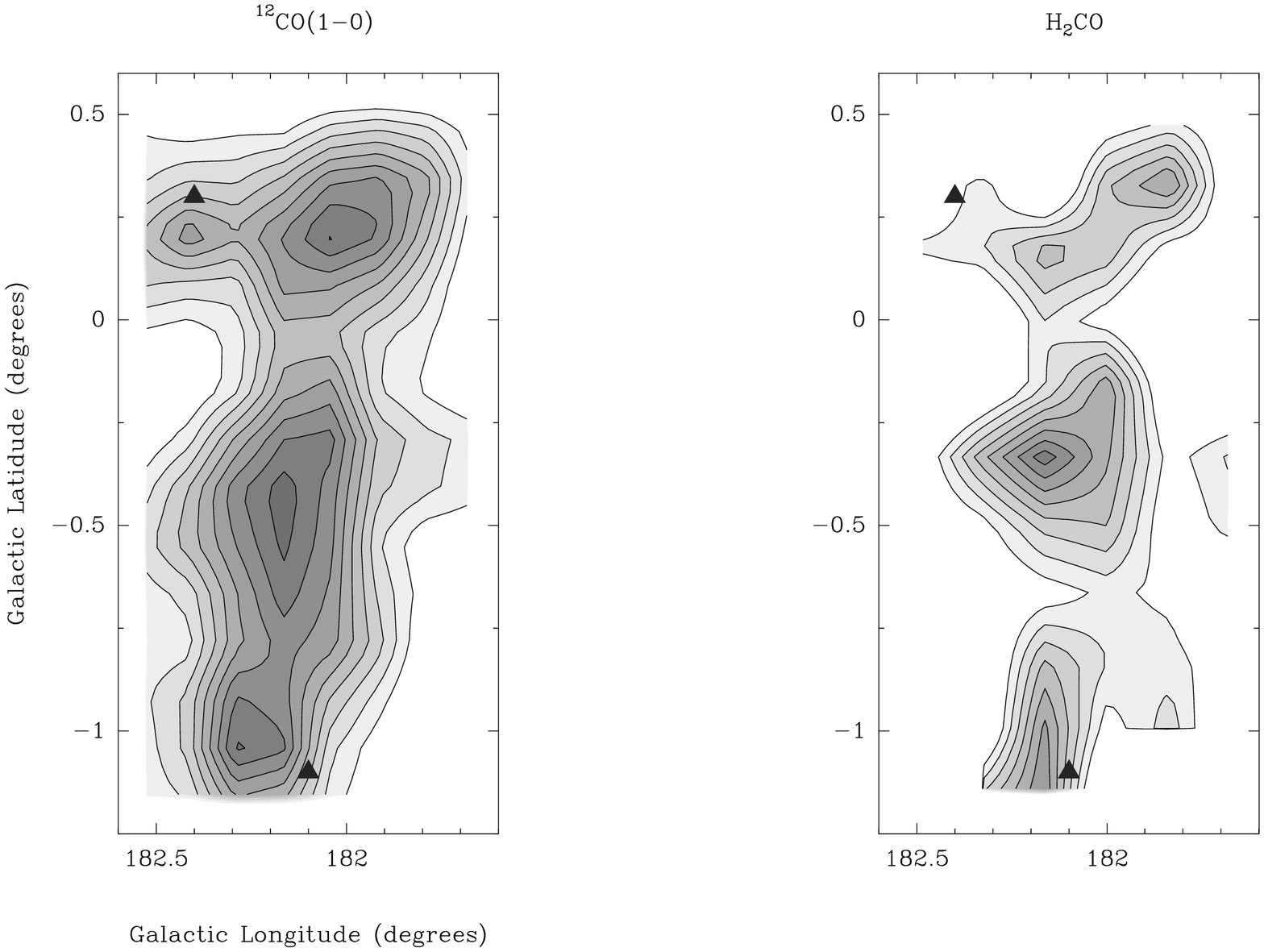}
\caption{a.(left panel) Velocity-integrated contour map of the $^{12}$CO emission for the first
region mapped in the direction $l = 182^{\circ}$, $b = 0^{\circ}$. b.(right panel) Velocity-
integrated \htco\ absorption for the same region. The blue triangles show the positions of
IRAS sources possibly associated with features in these regions. The velocity range extends
from  -16.0 km s$^{-1}$ to 6.0 km s$^{-1}$ in both panels. \label{fig:iras180}}
\end{figure*}

\begin{figure*}[ht!]
\plotone{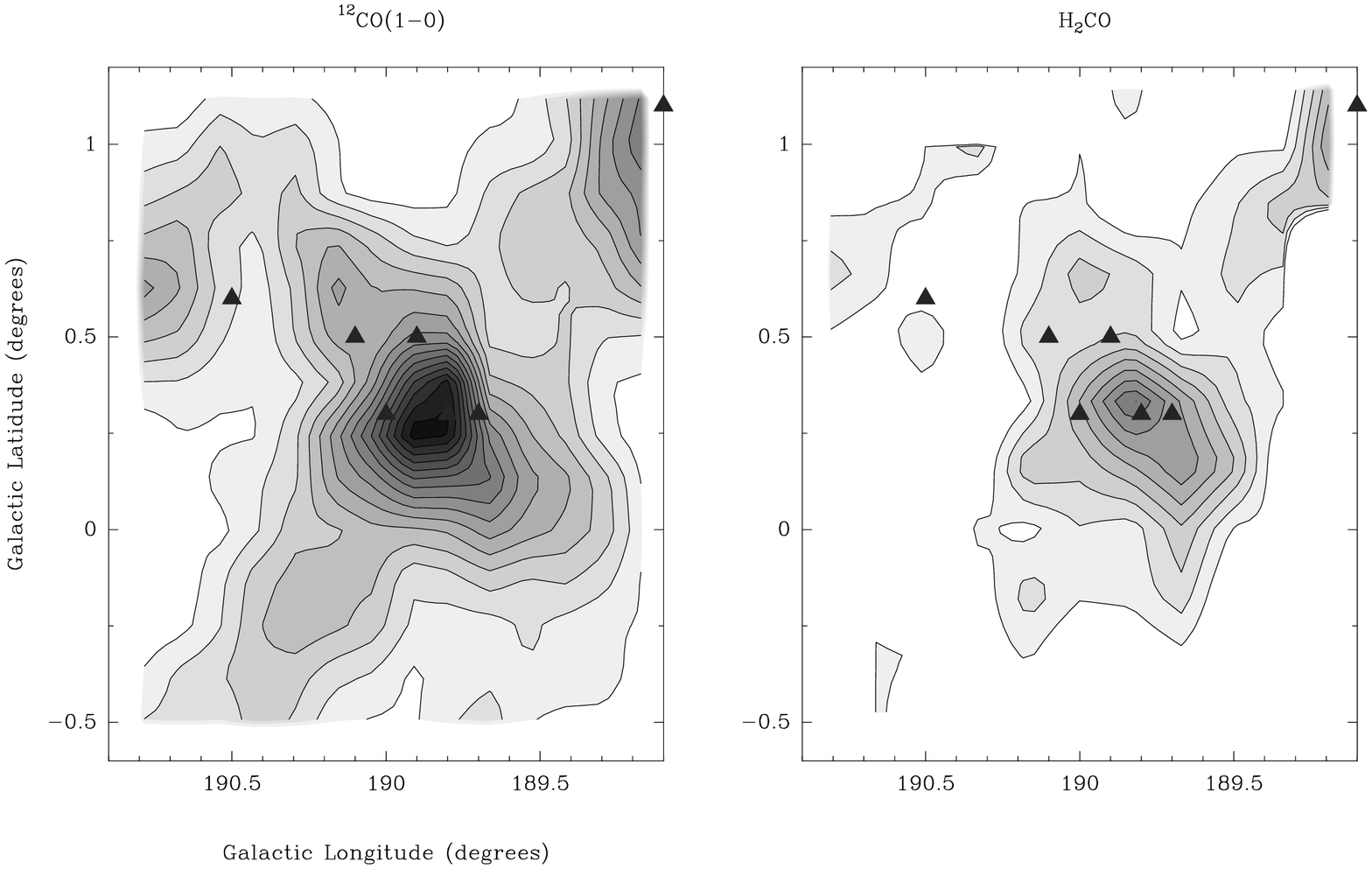}
\caption{a.(left panel) Velocity-integrated contour map of the $^{12}$CO emission for the second
region mapped in the direction $l = 190^{\circ}$, $b = 0^{\circ}$. b.(right panel) Velocity-
integrated \htco\ absorption for the same region. The blue triangles show the positions of
IRAS sources possibly associated with features in these regions. The velocity range extends from 
-3.0 km s$^{-1}$ to 16.0 km s$^{-1}$ in both panels. \label{fig:iras190}}
\end{figure*}

In Figures \ref{fig:iras180} and \ref{fig:iras190} we show contour plots of the \htco\ and
CO profile fluxes integrated over velocity for our two detailed fields. The IRAS sources from
Table \ref{table:IRAS} are indicated with the black triangles. We conclude that the correspondence
is only marginal, but the results suggest that the regions we have found are likely to be
located at distances of 1.5 - 3 kpc, probably in the Perseus arm of the Galaxy (2 -- 3 kpc
from the sun). In that case our $10'$ beam corresponds to a linear resolution of 4 - 9 pc.

\section{Discussion}
\label{sec:discussion}

The general correlation between the integrated line intensities of the CO(1-0) emission and the
\htco\ absorption suggests that the excitation characteristics of these two lines are similar.
In fact, both of these lines trace warm, dense molecular gas, as various calculations have shown
in many historical papers. For instance, \cite{hb97} show examples of emergent brightness
calculations for CO(1-0) in the Galactic GMC Lynds 1204 in the context of one popular model
(the large-velocity-gradient model).
As their Figure 1 shows, the observed brightness of this (usually optically-thick) line is
essentially proportional to the kinetic temperature of the gas as long as the density exceeds
the critical density of $\approx 10^3$ \pcmcub. It follows that any sensitivity-limited CO(1-0)
survey will therefore preferentially record the warm, dense regions of the ISM.

The details of \htco\ 6-cm line formation have been studied for example in several papers
by Evans and co-workers (\cite{mes87}; \cite{yle04}). The 6 cm absorption line arises through 
collisions to higher rotational levels followed by radiative
decay to the ground level, which becomes overpopulated owing to small differences in collision
cross sections. The line is therefore generally stronger if the collision rate is higher, so
this too favors warm, dense clouds. However, the absorption line is quenched at very high
collision rates, when the level populations approach a Boltzmann distribution,
and the 6 cm line then goes into emission. In a future paper on \htco\ mapping of the large
Galactic dust cloud Lynds 1204 \citep{r06} we will present model computations of \htco\ absorption
in more detail. For the moment we note that the 6 cm absorption is strongest in the density
range of $10^3 \lesssim n \lesssim 10^5$ \pcmcub. In our model the absorption is deepest
at $n \approx 10^{4.3}$ \pcmcub, first appearing at $T_K \gtrsim 10$ K, and saturating at
$T_K \approx 40$ K in the sense that the absorption hardly increases for higher temperatures.
This dependence on kinetic temperature is therefore somewhat different
than is the case for the CO(1-0) line; this is mostly a consequence of the fact that the \htco\
line remains optically thin, so it more closely reflects the specific excitation conditions.

One of the premises for doing the blind search for H$_2$CO towards the Galactic Anticenter
was the possibility of finding molecular gas in regions where CO emission had not previously
been detected. This could potentially reveal a component of the molecular gas characterized
by an excitation temperature lower than that corresponding to the energy level of the first rotational
transition of CO (or any other commonly observed molecule). Such an excitationally cold molecular
gas component would only be observable in absorption, either against a continuum background
source or, when using the anomalous absorption of H$_2$CO, against the CMB.
The fact that we do not see H$_2$CO absorption outside the regions where CO emission is
observed, suggests that either such cold molecular gas is not present, or, as we will argue,
the anomalous H$_2$CO absorption is not sensitive to excitationally cold molecular gas.

As shown by Townes \& Cheung (1969) and Evans et al. (1975),  the anomalous H$_2$CO
absorption is due to collisional pumping and most effective at densities between
$n_{\mathrm{H_2}} \approx 10^{3} - 10^{5}$ cm$^{-3}$.
Our observations of the 6-cm H$_2$CO line therefore seem to exclude the existence
of dense molecular gas outside the regions probed by CO emission. However, the effectiveness of
the collisional pumping at low temperatures has not been decisively determined (cf. Evans 1975;
Garrison et al. 1975), and we cannot exclude the possibility of very cold molecular gas outside
the CO emission regions, even if it meets the density requirement for inversion.
In Rodriguez et al. (2006) we found that the effectiveness of the collisional pumping leading to the
anomalous excitation of the 6-cm H$_2$CO transition is indeed temperature dependent, requiring
a relatively high temperature.
The combined requirement of a high temperature and density in order to render the anomalous
H$_2$CO absorption line observable means that whenever present, the H$_2$CO absorption
and CO emission lines will be spatially co-existing. Hence, the 6-cm H$_2$CO absorption seen
against the CMB is not a viable tracer of cold molecular gas. This is discussed in more detail in
Rodriguez et al. (2006).

\section{Conclusions}

\begin{itemize}

\item We have successfully detected anomalous CMB absorption by \htco-bearing dust clouds
at $\approx 10\%$ of our blind survey positions in the direction of the outer Galaxy. No emission
profiles were found.

\item Our success is likely due to the large improvement in sensitivity of our observations
over that of earlier surveys. This strongly indicates that the detection statistics will
improve further if even higher-sensitivity searches are carried out. Since the absorption
signal strength from more distant clouds will suffer from increasing beam dilution, future
searches ought to be done with larger radio telescopes.

\item H$_2$CO absorption and CO emission lines are spatially co-existing. All observed 
\htco\ absorption features were associated with know CO emission.

\item We have found a rough correspondence between the \htco\ and the $^{12}$CO(1-0) line
fluxes in detailed maps of two regions in our survey area. We have argued that both lines
generally trace warm, dense gas in the ISM, with the situation for CO(1-0) being somewhat
simpler owing to the fact that this line is usually optically thick.

\end{itemize}

\acknowledgements

We thank Prof.\ Roy Booth, director (ret.) of the radio observatory at Onsala, for generous
allocations of telescope time and for his warm hospitality during our several visits to the
observatory. We are also grateful to the observatory technical and
administrative staff for their capable assistance with our observing program. Dr.\ T.\ Dame
kindly provided a digital copy of relevant parts of the CfA CO(1-0) survey of the Galaxy.
We acknowledge the financial support of DGAPA, UNAM and CONACyT, in M\'exico, and the
\textit{Director's Discretionary Research Fund} at the Space Telescope Science Institute.


\appendix

\begin{deluxetable}{rrrrrc}
\tablewidth{0pt}
\tablecaption{Profile moments for each position of Fig. \ref{fig:firstmosaic}. The upper limits
correspond to $3 \sigma$. An ellipsis \ldots indicates no data were available; dashes ---
indicate data were available but no reliable fit could be made. Values marked with Y in
column 6 are plotted in Figure \ref{fig:intensityratio}. \label{table:mom182}}
\tablehead{
\colhead{Offset (\textit{l,b})} &
\colhead{$100 \times$ I(\htco)} & \colhead{$\langle V \rangle$} &
\colhead{I(CO)} & \colhead{$\langle V \rangle$} & \colhead{Include in} \\
\colhead{arcmin} &
\colhead{K \kmps} & \colhead{\kmps} &
\colhead{K \kmps} & \colhead{\kmps} & \colhead{Fig. \ref{fig:intensityratio} ?}
}
\startdata
30, -70 & \ldots  & \ldots  &  6.0 $\pm$ 1.7 & -9.6 $\pm$ 2.8\\
20, -70 & $<$ -9.3 &  ---  & 7.2 $\pm$ 1.7 & -10.6 $\pm$ 2.7 \\
10, -70 & $<$ -14.0  &  --- & 6.4 $\pm$ 0.8 & -11.9 $\pm$ 1.8 \\
        &   ---                 &   ---  &  6.7 $\pm$ 1.3 & -4.5 $\pm$ 1.0 \\
0, -70  & $<$ -14.0  & ---  &  4.5 $\pm$ 0.8 &-11.9 $\pm$ 2.6 \\
        &  ---                &        --- &  $<$ 3.9 & --- \\
-10, -70 & \ldots  & \ldots & $< $ 5.7 & --- \\
-20, -70 & \ldots  & \ldots & $<$ 6.0  & --- \\
\tableline
30, -60 &  $<$ -5.7  & ---  & $<6.9$ & --- \\
20, -60 & $<$ -6.9  & ---  & 10.0 $\pm$ 1.6 & -9.3 $\pm$ 1.6 \\ 
10, -60 & -6.7 $\pm$ 1.9  & -5.7 $\pm$ 3.2 & 7.3 $\pm$ 1.3 & -4.2 $\pm$ 0.8 & Y \\
              &  -4.5 $\pm$ 1.2 & -10.3 $\pm$ 3.3   &  12.2 $\pm$ 0.9 & -10.7 $\pm$ 0.9 & Y \\  
0, -60  & $<$ -3.0 & --- & $<$ 3.0 &--- \\
         &           ---              &     ---     &  6.7 $\pm$ 0.9 & -10.7 $\pm$ 1.8 \\
-10, -60 & $<$ -6.3  & --- & $<$ 6.0  & --- \\
-20, -60 & $<$ -6.9  & ---  & $<$ 6.0  & --- \\
\tableline
30, -50 &  $<$ -6.6  & ---  & 6.9 $\pm$ 1.7 & -9.1 $\pm$ 2.3 \\
20, -50 & $<$ -6.6  &  --- & 10.0 $\pm$ 1.7  & -9.4 $\pm$ 1.7  \\
10, -50 & -7.8 $\pm$ 2.1 & -3.2  $\pm$ 2.4 & 15.0 $\pm$ 1.0 &  -10.2 $\pm$ 0.8 & Y \\ 
                 &      ---            &         ---  &  $<$ 3.9 & --- \\  
                 & ---                  & ---          &   $<$ 3.3     & ---  \\
0, -50  &  -5.0 $\pm$ 1.0 & -9.7 $\pm$ 2.3 & 11.3 $\pm$ 1.6  & -8.9 $\pm$ 1.3 & Y \\
-10, -50 & $<$ -6.0  & ---  & 5.2 $\pm$ 1.2 & -10.6 $\pm$ 2.7 \\
           &    ---            &      ---             & $<$ 4.5 & --- \\
-20, -50 & $<$ -6.0 & ---  &  $<$ 6.9 & ---  \\
\tableline
30, -40 &  $<$ -6.3 & ---  & $<$ 6.0 & ---  \\
20, -40 &  $<$ -7.2 & --- & 8.0 $\pm$ 1.8 & -8.6 $\pm$ 2.1 \\
10, -40 & $<$ -4.2  & --- & 17.1 $\pm$ 1.2 & -10.5 $\pm$  0.8 \\
          &  ---             &       ---            &  $<$ 3.0 & --- \\
          & ---               & ---                  &  $<$ 2.5 & --- \\
0, -40 & $<$ -3.9  & ---  & 13.6 $\pm$ 1.2 & -10.2 $\pm$ 1.1 \\
         &    ---                    &    ---         &  $<$ 3.0 & --- \\
         &  ---                      &---              & $<$ 2.5 &   ---  \\
-10, -40 & $<$ -4.5 & ---  & 7.6 $\pm$ 1.5 & -10.5 $\pm$ 2.3 \\
              & ---            & --- &  $<$ 5.1      &  ---   \\
-20, -40 & $<$ -7.5  & ---  & $<$ 6.3 & --- \\
\tableline
30, -30 & $<$ -8.4  & ---  & 8.3 $\pm$ 1.8 & -8.3 $\pm$ 1.9 \\
20, -30 & $<$ -6.3  & ---  & 12.6 $\pm$ 2.0 & -8.7 $\pm$ 1.4 \\
10, -30 & -8.7 $\pm$ 2.5 & -8.8 $\pm$ 2.6  & 18.8 $\pm$ 1.0 & -10.6 $\pm$ 0.7 & Y \\
         &     ---                    &     ---                  &  $<$ 3.9 & --- \\
0, -30  & -8.6 $\pm$ 1.3  & -9.7 $\pm$ 1.3 &  14.3 $\pm$ 1.1 & -10.3 $\pm$ 0.9 & Y \\
            &  ---                       &        ---                 &  $<$ 4.2 & --- \\
-10, -30 & $<$ -9.0 &  --- & 5.8 $\pm$ 0.8 & -10.1 $\pm$ 1.7 \\
           &     ---           &       ---                    & $<$ 3.6 & --- \\
-20, -30 & -7.3 $\pm$ 2.3  & 0.7 $\pm$ 2.7  & $<$ 5.7 & --- \\
\tableline
30, -20 & $<$ -7.5 & --- & $<$ 6.3 & --- \\
20, -20 & -10.3 $\pm$ 1.9  & -9.4 $\pm$ 1.9 & 11.8 $\pm$ 1.8 & -8.6 $\pm$ 1.4 & Y \\
10, -20 & -15.8 $\pm$ 1.8 & -9.7 $\pm$ 0.9  & 19.1 $\pm$ 1.0 & -9.8 $\pm$ 0.6 & Y \\ 
         & ---                           &      ---               &  $<$ 3.3 & ---\\
0, -20 & -9.1 $\pm$ 1.3 & -9.7 $\pm$ 1.2  & 13.4 $\pm$ 1.1 & -10.4 $\pm$ 1.0 & Y \\
       &      ---                    &      ---                      &  4.6 $\pm$ 1.3 & 3.6 $\pm$ 1.2 \\
-10, -20 & $<$ -7.5 & --- &  4.6 $\pm$ 1.2 & -10.1 $\pm$ 3.1 \\
          &---                &       ---                  &  5.2 $\pm$ 1.5 & 3.3 $\pm$ 1.1 \\
-20, -20 & $<$ -7.5   & --- & $<$ 3.6 & --- \\
            & ---              &    ---                      &  $<$ 4.5 & --- \\
\tableline
30, -10 &  $<$ -7.5 & --- & $<$ 5.7 & --- \\
20, -10 & $<$ -4.8  & ---  & $<$ 6.0 & --- \\
10, -10 & -5.7 $\pm$ 1.3 & -9.6 $\pm$ 0.8  & 10.0 $\pm$ 1.0 & -9.9 $\pm$ 1.1 & Y \\
              &    ---          &      ---               & 4.0 $\pm$ 1.1 & 3.8 $\pm$ 1.2 \\ 
0, -10 & -11.3 $\pm$ 1.5  & -9.7 $\pm$ 0.8  & 9.4 $\pm$ 1.0 & -10.2 $\pm$ 1.2 & Y \\
          &  ---             &         ---            &   6.3 $\pm$ 1.2 & 2.8 $\pm$ 1.3 \\
-10, -10 & $<$ -6.0  & --- & $<$ 3.6 & --- \\
          &    ---           &         ---            &  $<$ 4.8 & --- \\
-20, -10 & $<$ -6.9  & --- & $<$ 4.5 & --- \\
         & ---              &      ---           &  $<$ 3.1 & --- \\
\tableline
30, 0 &  $<$ -6.0 & --- & $<$ 3.6 & --- \\ 
         &   ---            &        ---           &   $<$ 3.0 & --- \\
20, 0 & $<$ -6.6 & ---  & $<$ 3.3 & --- \\
            &   ---            &     ---            &  $<$ 2.7 & --- \\
10, 0 & $<$ -6.3 & ---  & 9.0 $\pm$ 1.1 & -10.4 $\pm$ 1.4 \\
           &   ---            &      ---           &  $<$ 4.2 & --- \\
0, 0  & $<$ -4.8 & --- & 8.1 $\pm$ 1.0 & -10.4 $\pm$ 1.5 \\
           & ---              &  ---               &   $<$ 3.9 & --- \\
-10, 0 & $<$ -5.7  & --- & $<$ 3.9 & ---  \\ 
           &  ---             &    ---    &  $<$ 3.8 & --- \\
-20, 0 &  $<$ -7.2 & ---  &  $<$ 3.6 & --- \\
           & ---              &     ---             &  $<$ 2.4 & --- \\  
\tableline
30, 10 & $<$ -7.5  & ---   & 7.0 $\pm$ 1.2 & 0.8 $\pm$ 1.3 \\
             & ---              &   ---            & $<$ 3.0 & --- \\ 
20, 10 & $<$ -5.7  & --- & 7.0 $\pm$ 1.4 & 1.4 $\pm$ 1.3 \\
     &   ---            &       ---              & $<$ 5.1 & --- \\
10, 10  & -8.3 $\pm$ 2.3  & 2.2 $\pm$ 0.8  & 7.3 $\pm$ 1.3 & 2.0 $\pm$ 1.3 & Y\\
     &  ---             &    ---                  &  8.5 $\pm$ 1.0 & -10.5 $\pm$ 1.5 \\
0, 10  & $<$ -4.5 & ---  & 7.2 $\pm$ 1.1 & 1.8 $\pm$ 1.3  \\
      &  ---             &   ---                  &  8.5 $\pm$ 1.0 & -9.7 $\pm$ 1.3  \\
-10, 10 & $<$ -6.6 & --- & 6.7 $\pm$ 1.2 & 2.6 $\pm$ 1.3 \\
       &     ---          &   ---  &  6.3 $\pm$ 0.9 & -9.9 $\pm$ 1.7 \\
-20, 10 & $<$ -8.1 & --- &   $<$ 4.2 & --- \\ 
     &  ---             &    ---              & $<$ 3.0 & --- \\ 
\tableline
30, 20 & $<$ -8.2  & ---  & $<$ 3.9 & --- \\
   &   ---            &    ---            &   3.7 $\pm$ 1.0 & -9.7 $\pm$ 3.2 \\
20, 20 & $<$ -6.6   & ---  & 6.9 $\pm$ 1.4 & 1.2 $\pm$ 1.3 \\
    &   ---            &      ---              &  $<$ 3.0 & --- \\ 
10, 20 & $<$ -6.0   & ---  & 7.4 $\pm$ 1.2 & 1.9 $\pm$ 1.3 \\ 
    &   ---            &     ---            &  4.3 $\pm$ 1.2 & -9.1 $\pm$ 2.8 \\
0, 20 & -7.3 $\pm$ 1.6 & 1.5 $\pm$ 0.5  & 10.3 $\pm$ 1.2 & 2.3 $\pm$ 1.3 & Y \\
            &   ---            &    ---             &  6.7 $\pm$ 1.2 & -8.7 $\pm$ 1.7 \\  
-10, 20 & -7.1  $\pm$ 1.0    & 2.1 $\pm$ 0.4 & 10.1 $\pm$ 1.2 & 2.5 $\pm$ 1.3 & Y \\
               &  $<$ -4.5  &  --- &  7.7 $\pm$ 1.2 & -8.8 $\pm$ 1.6 \\ 
-20, 20  & $<$ -5.7 & --- &  $<$ 4.5 & --- \\ 
           &    ---           &      ---            & 3.6 $\pm$ 1.1 & -8.9 $\pm$ 3.2 \\
\tableline
30, 30 & $<$ -6.9  & ---  & $<$ 3.9 & --- \\ 
              &     ---         &     ---        & 3.7 $\pm$ 1.0 & -9.4 $\pm$ 3.0 \\
20, 30 & $<$ -9.0 & ---  & $<$ 3.0 & --- \\   
          &  ---             &     ---      & $<$ 3.6 & --- \\
10, 30 &  $<$ -9.9 & ---  & $<$ 4.5 & --- \\
           &  ---             &   ---             & $<$ 3.3 & --- \\
0, 30  & $<$ -3.9  & ---  & 4.5 $\pm$ 1.0 & -8.6 $\pm$ 2.1 \\
        &  ---             &      ---          &  $<$ 3.7 & --- \\
-10, 30 & $<$ -7.2 & --- & $<$ 3.9 & --- \\ 
     &  ---             &       ---        &   5.4 $\pm$ 1.0 & -8.7 $\pm$ 1.9 \\
-20, 30 & $<$ -5.7 & ---  & $<$ 3.3 & --- \\
           &  ---             &    ---       &   $<$ 3.0 & --- \\
\enddata
\end{deluxetable}


\begin{deluxetable}{rrrrrc}
\tablewidth{0pt}
\tablecaption{Moments for each survey position of Fig. \ref{fig:secondmosaic}. The upper limits
correspond to $3 \sigma$. An ellipsis \ldots indicates no data were available; dashes ---
indicate data were available but no reliable fit could be made. Values marked with a Y in
column 1 are plotted in Figure \ref{fig:intensityratio}. \label{table:mom190}}
\tablehead{
\colhead{Offset (\textit{l,b})} &
\colhead{$100 \times$ I(\htco)} & \colhead{$\langle V \rangle$} &
\colhead{I(CO)} & \colhead{$\langle V \rangle$} & \colhead{Include in} \\
\colhead{arcmin} &
\colhead{K \kmps} & \colhead{\kmps} &
\colhead{K \kmps} & \colhead{\kmps} & \colhead{Fig. \ref{fig:intensityratio} ?}
}
\startdata
50 -30 & \ldots & \ldots &  5.8 $\pm$ 0.9 & 6.6 $\pm$5.6 \\
40 -30 & $<$ -6.0& --- & 6.2 $\pm$ 0.7   &  8.5 $\pm$ 6.2 \\
               &  ---   & ---  & 6.3 $\pm$ 0.5 &  0.2 $\pm$ 6.0 \\
30 -30 & $<$ -8.1 & ---  &   7.1 $\pm$ 0.9 & 9.0 $\pm$ 6.0 \\ 
               &   ---            &    ---   & $<$ 3.3   & --- \\
20 -30 & $<$ -5.4   & ---  & 7.1 $\pm$ 0.7 &  11.6 $\pm$ 5.0 \\
               &       ---        &         ---  &  9.1 $\pm$ 0.7  & 5.4 $\pm$ 3.4   \\
               & --- & --- & $<$ 1.5 & --- \\
10 -30 &  $<$ -5.1 & --- & 7.4 $\pm$ 0.9 & 6.7 $\pm$ 5.2  \\
0 -30 &  $<$ -3.6  & ---  & 5.8 $\pm$ 0.8 & 5.1 $\pm$ 4.6  \\
-10 -30 & $<$ -5.7 & --- &  6.0 $\pm$ 1.0 & 6.3 $\pm$ 6.3  \\
-20 -30 &  $<$ -5.4 & --- & $<$ 2.3  & --- \\
               &         ---      &--- &  9.5 $\pm$ 0.5 & 4.8 $\pm$ 2.2 \\  
               & --- & --- &  $<$ 2.3 & --- \\
-30 -30 & $<$ -5.1  & --- &  7.7 $\pm$ 1.1 & 7.4 $\pm$ 6.3   \\
-40 -30 & $<$ -6.9  & --- &  $<$ 2.9 & --- \\
-50 -30 & \ldots  & \ldots & $<$ 3.0  & --- \\
\tableline
50 -20 & \ldots  & \ldots  &  $<$ 2.4 & --- \\
            & \ldots  & \ldots  &  $<$ 3.0 & --- \\
40 -20 & $<$ -5.7 & ---  & 6.0 $\pm$ 0.8 & 7.0 $\pm$ 6.3 \\
               &    ---    &---   & 5.3 $\pm$ 0.5   & -0.5 $\pm$ 5.0    \\
30 -20 & $<$ -5.7 & ---&  10.1 $\pm$ 0.7 & 9.4 $\pm$ 3.9  \\
               &    ---           &     --- & $<$ 3.9 &  --- \\
20 -20 &  $<$ -5.4  & --- &  8.6 $\pm$ 0.7 & 11.2 $\pm$ 4.0 \\
               &      ---        &     --- & 7.9 $\pm$ 0.7 &  5.4 $\pm$ 3.7   \\
               & ---           & ---       & $<$ 1.5  &     ---   \\
10 -20 &  $<$ -5.7 & --- & 12.1 $\pm$ 1.0 &  9.1 $\pm$ 4.2 \\
0 -20 &  $<$ -3.3 & --- &  10.2 $\pm$ 1.0 & 6.7 $\pm$ 3.7 \\
-10 -20 & $<$ -6.0   & ---  & 5.9 $\pm$ 0.9   & 5.6 $\pm$ 5.2   \\
-20 -20 & $<$ -5.4 & --- & 9.8 $\pm$ 1.0  & 6.7 $\pm$ 4.0  \\
-30 -20 & $<$ -7.2   & --- & 10.3 $\pm$ 0.8 & 7.1 $\pm$ 3.1  \\
-40 -20 &  $<$ -4.8 & --- &  7.2 $\pm$ 1.0 & 9.8 $\pm$ 6.0 \\
               &    ---  &      --- &  $<$ 3.2 & ---  \\
-50 -20 & \ldots  & \ldots & $<$ 4.0 & ---  \\
\tableline
50 -10 & \ldots  & \ldots  &  10.0  $\pm$ 0.9 & 9.3 $\pm$ 5.2   \\
40 -10 & $<$ -5.4   & ---  & $<$ 3.3  & ---   \\
30 -10 & $<$ -5.7 & ---  & 9.5 $\pm$ 0.8 & 7.9 $\pm$ 4.1   \\
20 -10 & $<$ -5.1  & ---  & 16.5 $\pm$ 0.9   & 6.5 $\pm$ 2.3   \\
10 -10 & $<$ -7.8  & --- & 20.3 $\pm$ 1.0  &  8.7 $\pm$ 2.7 \\
0 -10  & $<$ -3.0 & --- & 19.2 $\pm$ 0.9   & 8.1 $\pm$ 2.4   \\
-10 -10 & $<$ -6.6 & --- & 10.9 $\pm$ 1.4  & 7.0 $\pm$ 5.4   \\
-20 -10 & -10.7 $\pm$ 2.9  & 6.7 $\pm$ 6.7 & 16.9 $\pm$ 1.0 &  7.7 $\pm$ 2.8  & Y \\
-30 -10 & $<$ -5.4  & --- &  10.5 $\pm$ 0.7  & 8.4 $\pm$ 3.8   \\
             &   ---        & --- & $<$ 3.3 & --- \\
-40 -10 & $<$ -6.9 & --- & 11.2 $\pm$ 0.9  &  9.2 $\pm$ 4.4  \\
             & --- & --- & $<$ 4.2 & --- \\
-50 -10 & \ldots  & \ldots & 6.4 $\pm$ 1.0  & 8.7 $\pm$ 7.6  \\
\tableline
50 0 & \ldots  & \ldots & 8.7 $\pm$ 1.1   &  7.8 $\pm$ 5.9 \\
40 0 & $<$ -6.0  & ---  &  6.1 $\pm$ 0.8  & 6.6 $\pm$ 5.1   \\
30 0 & $<$ -5.4 & --- & 5.3 $\pm$ 1.1  & 7.0 $\pm$ 8.9  \\
20 0 & $<$ -5.1 & --- &  9.9 $\pm$ 1.2  & 6.8 $\pm$ 4.8   \\
10 0 & $<$ -6.6   & ---  & 17.8 $\pm$ 0.8  & 8.1 $\pm$ 2.3   \\
0 0 & -8.2 $\pm$ 1.3  & 7.9 $\pm$ 4.8 & 20.9 $\pm$ 1.0  & 8.2 $\pm$ 2.4 & Y  \\
-10 0 & -5.8 $\pm$ 1.6 & 7.8 $\pm$ 8.8 & 21.2 $\pm$ 0.9   & 7.6 $\pm$ 2.0  & Y  \\
-20 0 & -14.8 $\pm$ 2.4  & 8.5 $\pm$ 5.4 & 28.2 $\pm$ 1.0 & 7.6 $\pm$ 1.8  & Y \\
-30 0 &  $<$ -5.4 & --- &  9.1 $\pm$ 0.7   &  1.2 $\pm$ 3.1  \\
               &       ---        &      ---        & 15.4 $\pm$ 0.9 & 8.7 $\pm$ 3.1 \\ 
-40 0 &  $<$ -6.6 & ---   &  9.1 $\pm$ 0.6 & 1.2  $\pm$ 1.7 \\
               &   ---            &      ---         &   12.0 $\pm$ 0.8 & 8.9 $\pm$ 3.9 \\
-50 0 & \ldots & \ldots &  $<$ 4.5 & --- \\
               &   \ldots      &  \ldots   &   $<$ 2.6  &  ---   \\
\tableline
50 10 & \ldots &  \ldots &  5.4 $\pm$ 1.1 & 6.2 $\pm$ 7.6   \\
40 10 &  $<$ -6.9  & ---  & $<$ 3.6 & ---  \\
30 10 &  $<$ -6.0 & --- &  7.7 $\pm$ 1.1 & 5.5 $\pm$ 4.8   \\
20 10 & $<$ -4.8   & --- & 9.7 $\pm$ 1.1  & 6.2 $\pm$ 4.3   \\
10 10 &  -14.8 $\pm$ 2.3 & 6.3 $\pm$ 4.7 & 25.7 $\pm$ 0.9   &  7.7 $\pm$ 1.7 & Y \\
0 10 & -11.0 $\pm$ 1.7  & 7.0 $\pm$ 5.0 & 31.3 $\pm$ 1.4  & 7.9 $\pm$ 2.1  &Y \\
-10 10 & -16.5 $\pm$ 2.2  & 8.3 $\pm$ 5.2 & 40.5 $\pm$ 0.8  & 7.5 $\pm$ 1.2 &Y  \\
-20 10 & -22.2 $\pm$ 1.9 & 7.9 $\pm$ 3.2 & 38.9 $\pm$ 0.8  &  7.4 $\pm$ 1.1 &Y \\
-30 10 & $<$ -6.6 & --- & 18.2 $\pm$ 0.6 & 8.5 $\pm$ 2.0 \\
               &        ---       & ---   & 9.6 $\pm$ 0.5   &  0.7 $\pm$ 1.5 \\
-40 10 & $<$ -5.7  & --- & 14.1 $\pm$ 0.6 & 7.8 $\pm$ 2.5 \\ 
               &---               &    --- &  9.1 $\pm$ 0.5  &   1.4 $\pm$ 1.7   \\
-50 10 & \ldots & \ldots & 5.7 $\pm$ 1.1   & 4.1 $\pm$ 5.3   \\
\tableline
50 20 &  \ldots & \ldots &  15.4 $\pm$ 1.0 & 6.3 $\pm$ 2.5   \\
40 20 & $<$ -6.0  & ---  &  9.9 $\pm$ 1.0 & 6.1 $\pm$ 3.4  \\
30 20 & $<$ -5.1 & ---  & 7.7 $\pm$ 0.9   & 5.7 $\pm$ 3.8   \\
20 20 & $<$ -6.9  & --- & 7.0 $\pm$ 0.9   & 7.0 $\pm$ 5.4  \\
10 20 & $<$ -6.6  & ---  & 15.0 $\pm$ 0.8  & 7.6 $\pm$ 2.6 \\
0 20 & -17.3 $\pm$ 1.7  & 8.2 $\pm$ 3.8 & 29.4 $\pm$ 0.9 & 7.6 $\pm$ 1.5  &Y \\
-10 20 &  -29.4 $\pm$ 2.0 & 8.3 $\pm$ 2.7 &  50.2 $\pm$ 0.6 & 7.9 $\pm$ 0.9  &Y \\
-20 20 & -19.8 $\pm$ 2.0  & 8.0 $\pm$ 3.8 & 19.7 $\pm$ 0.5  & 8.3 $\pm$ 1.4  &Y \\
-30 20 & $<$ -7.5 & --- & 17.7 $\pm$ 0.7  & 8.7 $\pm$ 2.0  \\
-40 20 & $<$ -6.9  & --- & 10.5 $\pm$ 0.6 & 9.6 $\pm$ 3.5 \\
               &       ---        &      ---                  & $<$ 4.8  & ---  \\ 
-50 20 &  \ldots & \ldots & 5.2 $\pm$ 0.9  & 4.2 $\pm$ 4.9   \\
\tableline
50 30 & $<$ -11.4  & ---  & 14.6 $\pm$ 1.2   &  7.1 $\pm$ 3.7 \\
40 30 & $<$ -5.7  & --- & 17.6 $\pm$ 1.3   & 6.2 $\pm$ 2.3   \\
30 30 & -9.3 $\pm$ 1.8 & 7.2 $\pm$ 6.4 &  11.4 $\pm$ 1.0 & 6.7 $\pm$ 3.5 & Y   \\
20 30 & $<$ -4.2  & --- &  7.0 $\pm$ 0.8 & 6.5 $\pm$ 4.2   \\
10 30 & $<$ -5.7  & ---  &  24.6 $\pm$ 1.1 &  8.0 $\pm$ 2.3 \\
0 30  & -10.5 $\pm$ 1.3 & 6.4 $\pm$ 6.3 & 22.9 $\pm$ 0.9   &  8.1 $\pm$ 1.9  &Y\\
-10 30 & -9.6  $\pm$ 2.2 & 4.7 $\pm$ 5.4 & 30.5 $\pm$ 0.7  & 7.2 $\pm$ 1.1 & Y \\
-20 30 & $<$ -5.4 & --- &  12.9 $\pm$ 0.6 &  7.3 $\pm$ 1.9  \\
-30 30  & $<$ -4.5 & --- &  $<$ 4.9 & --- \\
               &     ---          &     --- &  6.8 $\pm$ 0.5 & 9.4 $\pm$ 4.1 \\
-40 30 & $<$ -6.6  & --- & 6.9 $\pm$ 0.5 & 2.8 $\pm$ 4.8 \\
               &     ---          &      ---&   8.1 $\pm$ 0.5   &  9.9 $\pm$ 4.0  \\
-50 30 & \ldots  & \ldots &  9.4 $\pm$ 0.9 & 4.5 $\pm$ 2.8   \\
\tableline
50 40 & $<$ -12.0  & ---  & 18.9 $\pm$ 0.8   & 6.7 $\pm$ 1.9   \\
40 40 &  $<$ -6.3 & --- & 22.4 $\pm$ 1.0  & 7.3 $\pm$ 2.0 \\
30 40 & $<$ -6.6 & --- & 13.5 $\pm$ 0.9  & 6.7 $\pm$ 2.8   \\
20 40 & $<$ -6.6  & --- & 8.1 $\pm$ 1.0  & 6.7 $\pm$ 4.8   \\
10 40 & $<$ -6.3  & --- & 28.4 $\pm$ 1.0  & 7.6 $\pm$ 1.7  \\
0 40 &  -12.8 $\pm$ 1.4 &  7.2 $\pm$ 3.6 & 22.9 $\pm$ 1.0  & 7.8 $\pm$ 2.0 &Y  \\
-10 40 & -11.0 $\pm$ 2.2  & 7.4 $\pm$ 5.7 & 22.3 $\pm$ 0.8  & 6.9 $\pm$ 1.7 &Y  \\
-20 40 & $<$ -7.5  & --- & 12.5 $\pm$ 0.8  & 6.4 $\pm$ 2.4   \\
-30 40 & $<$ -7.8 & --- & 15.5 $\pm$ 0.8  &  5.7 $\pm$ 1.7 \\
-40 40 & $<$ -6.0  & --- & 10.2 $\pm$ 0.4 & 3.8 $\pm$ 1.4 \\
               &        ---       &          ---   &    $<$ 3.8 & --- \\
-50 40 &  \ldots & \ldots & 21.8 $\pm$ 1.0   & 4.0 $\pm$ 1.4   \\
\tableline
50 50 & $<$ -8.4  & ---  & 6.0  $\pm$ 1.1 & 5.6 $\pm$ 6.0   \\
40 50 & $<$ -8.1 & --- &  14.7 $\pm$ 1.2  & 6.0 $\pm$ 2.8 \\
30 50 & $<$ -6.6 & --- & 16.9 $\pm$ 0.9 &  6.6 $\pm$ 2.3  \\
20 50 & $<$ -7.5  &  --- & 12.9 $\pm$ 1.1  & 7.2 $\pm$ 3.8  \\
10 50 &  $<$ -7.8 & --- & 7.4 $\pm$ 1.1 & 8.1 $\pm$ 6.9   \\
0 50 &  $<$ -3.6 & ---  & $<$ 4.0 & --- \\
-10 50 & $<$ -6.0  & --- & $<$ 4.5  & ---   \\
-20 50 & $<$ -5.7  & --- & 11.1 $\pm$ 0.8   & 5.4 $\pm$ 2.4   \\
-30 50 &  $<$ -6.6 & --- & 15.2 $\pm$ 1.2    & 4.4 $\pm$ 2.2  \\
-40 50 & -13.6 $\pm$ 1.4  & 6.7 $\pm$ 4.4  & 17.2  $\pm$ 1.0 &  4.6 $\pm$ 1.8 &Y  \\
-50 50 & -13.1 $\pm$ 2.4 & 7.0 $\pm$ 4.3  &  31.2 $\pm$ 0.8 & 6.0 $\pm$ 1.1  &Y \\
\tableline
50 60 & \ldots  & \ldots &   5.7 $\pm$ 1.0 & 4.8 $\pm$ 4.9   \\
40 60 &  $<$ -4.2 & ---  & 8.3 $\pm$ 1.2  & 4.4 $\pm$ 3.8   \\
30 60 &  $<$ -5.7 & --- & 14.5 $\pm$ 0.9 & 5.9 $\pm$ 2.2   \\
20 60 & -11.9 $\pm$ 1.8 & 7.9 $\pm$ 5.1 & 8.7 $\pm$ 1.0   & 7.9 $\pm$ 3.8 &Y  \\
           & ---      & ---       & $<$ 2.2 & --- \\
10 60 & $<$ -6.9  & ---   & $<$ 1.5 & ---   \\
          &   ---           & --- & $<$ 4.2  & --- \\
          &   ---           & --- & $<$  2.5 & --- \\
0 60 & $<$ -4.2  & --- & $<$ 3.6  & --- \\
-10 60 & $<$ -5.4  & --- & $<$ 4.5 & ---    \\
-20 60 & $<$ -8.4  & ---  & 7.3 $\pm$ 0.9   & 4.3 $\pm$ 3.2   \\
-30 60 &  $<$ -6.0 & ---  &  7.1 $\pm$ 0.7 &  5.2 $\pm$ 3.3  \\
-40 60 & $<$ -7.8 & --- & 11.7 $\pm$ 1.0  & 5.0 $\pm$ 2.6   \\
-50 60 & -23.2 $\pm$ 2.7 & 5.8 $\pm$ 3.3 & 37.6 $\pm$ 0.9  & 6.2 $\pm$ 1.0 &Y  \\
\tableline
50 70 &  \ldots & \ldots  & $<$ 4.6 & ---  \\
40 70 &  \ldots & \ldots &  $<$ 3.0 & ---   \\
30 70 & \ldots  & \ldots &  8.3 $\pm$ 1.1 & 7.2 $\pm$ 5.7  \\
20 70 & \ldots  & \ldots & 7.5 $\pm$ 0.9  & 9.8 $\pm$ 5.0   \\
          & \ldots  & \ldots & $<$ 2.4 & ---  \\
10 70 & \ldots & \ldots & $<$ 5.1  & ---   \\
0 70 & $<$ -6.0  & ---  & $<$ 3.9  & ---   \\
-10 70 & $<$ -7.8 & --- & $<$ 4.6   & ---  \\
-20 70 & $<$ -6.9   & --- & $<$ 4.6 & ---  \\
-30 70 & $<$ -7.5  & --- & $<$ 2.6  & --- \\
            & --- & --- & $<$ 4.8 & --- \\
-40 70 & $<$  -6.6 & --- &  7.4 $\pm$ 0.7 & 2.1 $\pm$ 5.7 \\
               &  ---             & --- &  7.1 $\pm$ 0.7    &  8.9 $\pm$ 5.8   \\
-50 70 &  -18.5 $\pm$ 2.7  & 4.6 $\pm$ 3.5 & 34.7 $\pm$ 0.9  & 4.6  $\pm$ 1.0 &Y \\
\enddata
\end{deluxetable}

\end{document}